\documentclass{jpsj2}
\newcommand{\be}{\begin{equation}}
\newcommand{\ee}{\end{equation}}
\renewcommand{\l}{b}
\newcommand{\pl}{P}
\newcommand{\z}{z}

\newcommand{\Z}{{\mathcal{Z}}}
\newcommand{\C}{{\mathcal{C}}}
\newcommand{\Pl}{{\mathcal{P}}}
\newcommand{\Ss}{{\mathcal{S}}}

\title{Error counting in a quantum error-correcting code
and the ground-state energy of a spin glass}

\author{Hidetoshi \textsc{Nishimori}$^{1}$ and Peter
\textsc{Sollich}$^{2}$}

\inst{$^{1}$Department of Physics, Tokyo Institute of Technology,
Oh-okayama, Meguro-ku, Tokyo 152-8551 \\
$^{2}$Department of Mathematics, King's College London, Strand, London
WC2R 2LS, United Kingdom}

\abst{Upper and lower bounds are given for the number of equivalence
classes
of error patterns in the toric code for quantum memory. The results
are used to derive a lower bound on the ground-state energy of the
$\pm J$ Ising spin glass model on the square lattice with symmetric
and asymmetric bond distributions.  This is a highly non-trivial
example in which insights from quantum information lead directly to an
explicit result on a physical quantity in the statistical mechanics of
disordered systems.}

\kword{quantum information, quantum memory, toric code, spin glass,
$+/-J$ model, ground-state energy}

\begin{document}
\maketitle

\section{Introduction}

In a recent paper, Dennis, Kitaev, Landall and Preskill (DKLP)
\cite{DKLP} showed that error correction in the toric code for stable
storage of quantum information is closely related to the phase
transition between ferromagnetic and paramagnetic phases in the
two-dimensional $\pm J$ Ising model of spin glasses. This is
remarkable in that it paves a way for transferring results, concepts
and insights between the two at first sight unrelated domains of
quantum error correction and spin glasses, and provides the motivation
for the present work.

An important aspect of the toric code is that it has a finite error
threshold;
this is the critical value of error probability per qubit beyond which
it is
impossible to correct errors in the thermodynamic (large system-size)
limit.
Thus, for error rates larger than the threshold,
the encoded quantum information is lost. It is therefore
necessary to estimate the precise value of the error threshold for,
e.g., the design of hardware of quantum memory.
DKLP have shown that this error threshold is equal to the probability
of antiferromagnetic bonds at the multicritical point in the phase
diagram
of the $\pm J$ Ising model on the square lattice.

We do not discuss the value of error threshold itself in the present
paper
\cite{DKLP,MKN,NK}.
We instead strive to clarify a closely related problem of
the structure of error patterns underlying
a characteristic feature of the toric code, {\em degeneracy}, by
deriving
bounds on the number of equivalent error patterns.
These bounds are shown to be related to the information-theoretical
entropy of the
distribution of frustrated plaquettes in the $\pm J$ Ising model.
It will also be shown that a lower bound on the ground-state energy
of the $\pm J$ Ising model can be obtained from the bounds derived for
the toric code.

We outline the link between spin glasses and quantum error correcting
codes in \S\ref{review}. Errors are detected by measurement of a
syndrome, from which the underlying error pattern is to be inferred
and then corrected. Due to the nature of quantum encoding in the toric
code,
however, many error patterns are equivalent (degenerate), and it is
sufficient
to infer the equivalence class of the true error pattern, not the
precise
pattern itself. In
\S\ref{overall} we count the number of syndromes $D$ and the
total number of equivalence classes $C$, which is an easy task. The
hard part is counting the number of equivalence classes $C(p)$
containing error patterns with a given fraction $p$ of errors. We
first ignore the issue of degeneracy, and count error patterns instead
of equivalence classes, from which upper bounds on $C(p)$ are derived
in \S\ref{upper}. Lower bounds are discussed in
\S\S\ref{lower} and \ref{lower2}.
These bounds are summarized in \S\ref{bounds_s}.
One of the lower bounds involves the ground-state
energy of the $\pm J$ Ising model, and this fact is used in
\S\ref{ground_state} to derive a lower bound on the latter
ground-state energy. The results are summarized and discussed in
\S\ref{conclusion}.

\section{Toric code and spin glass}
\label{review}

It is useful to first sketch the connection between toric quantum
codes
and two-dimensional spin glasses, following DKLP~\cite{DKLP}, since
most of
the readers may be unfamiliar with this relatively new
interdisciplinary field.

\subsection{Encoding in the toric code}

Consider a square lattice with
$N=L^2$ sites and $n=2N$ bonds labelled by $\l=(ij)$ and connecting
nearest-neighbour
sites $i$ and $j$; toroidal boundary conditions are used so that bonds
on a boundary connect to the opposite side of the lattice. The toric
code comprises $n$ quantum (spin-$1/2$) spins located on the bonds of
the lattice. We call the local Pauli operators $X_\l$, $Y_\l$ and
$Z_\l$. In the $Z$-basis of the state space, each basis vector
$|\z\rangle = |\{\z_\l\}\rangle$ is specified by the $z$-components
$\z_\l=\pm 1$ of all spins. The Pauli operators then act as $Z_\l
|\ldots \z_\l \ldots\rangle = \z_\l |\ldots \z_\l \ldots\rangle$ and
$X_\l |\ldots \z_\l \ldots\rangle = |\ldots -\!\z_\l \ldots\rangle$ so
that $X_\l$ effectively just flips the spin at $\l$; also, $Y_\l =
iX_\l Z_\l$.

\begin{figure}[t]
\begin{center}
\includegraphics[width=0.25\linewidth]{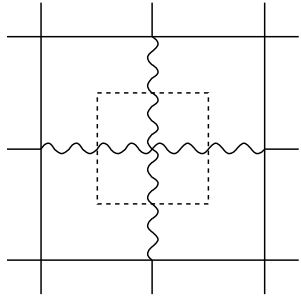}
\end{center}
\caption{Example of a cycle. Full lines represent bonds with $z_b=1$
and wavy lines are for $z_b=-1$. Dual bonds to $z_b=-1$  form a closed
loop (shown dotted).}
\label{fig:cycles}
\end{figure}

Quantum states are vulnerable to decoherence, and robustness should be
introduced to protect quantum information.
Quantum error correction is a powerful method for this purpose,
in which one encodes quantum information by mapping it to another
(redundant) set of quantum states.
Specifically, in the toric code, one maps the state
space of two logical qubits that are to be encoded onto a
$2^2=4$-dimensional subspace (the ``code space'') of the spin system's
$2^n$-dimensional state space. This mapping is redundant because $n$
qubits are used to represent 2 qubits.
A basis $|\psi_0\rangle, \ldots,
|\psi_3\rangle$ for the code space can be defined as
\begin{equation}
|\psi_i\rangle \propto \sum_{\z\in \C_i} |\z\rangle,
\label{codespace}
\end{equation}
where $\C_i$ denotes a class
of states ({\em equivalence class}) to be defined shortly.
The sum here runs over states $|\z\rangle$ which form {\em cycles}
within an
equivalence class.  This means that out of the four bonds $\l(\pl)$
around each
lattice plaquette $\pl$ an even number are negative ($\z_\l=-1$). A
cycle $|\z\rangle$ is thus an eigenstate with eigenvalue 1 of all the
operators $Z_\pl=\prod_{\l(\pl)} Z_\l$. Geometrically, $|\z\rangle$ is
a cycle if the duals to its negative bonds form closed loops on the
dual lattice (see Fig.~\ref{fig:cycles}).

\subsection{Equivalence class}

\begin{figure}[htb]
\begin{center}
\includegraphics[width=0.5\linewidth]{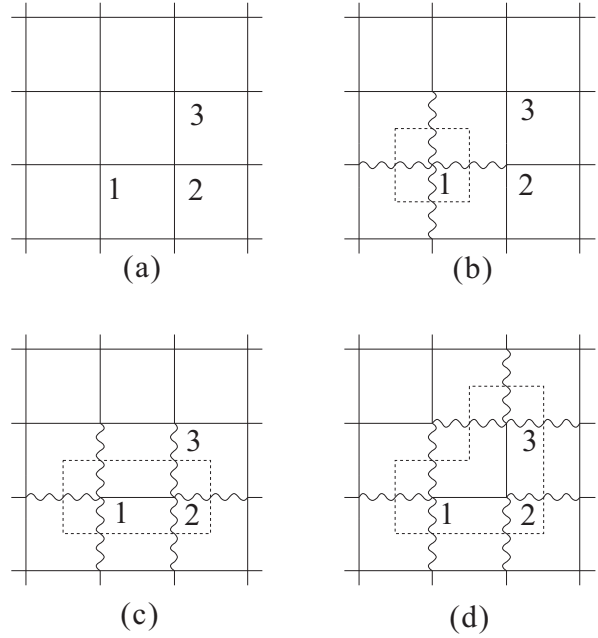}
\end{center}
\caption{Four equivalent cycles. Full, wavy and dotted lines have
the same meaning as in Fig.~\ref{fig:cycles}.
Application of $X_1$ to (a) gives (b); similarly, (c) and (d) are
obtained from (a) by applying $X_2 X_1$ and $X_3 X_2 X_1$,
respectively.
\label{fig:cycles2}
}
\end{figure}

\begin{figure}[htb]
\begin{center}
\includegraphics[width=0.5\linewidth]{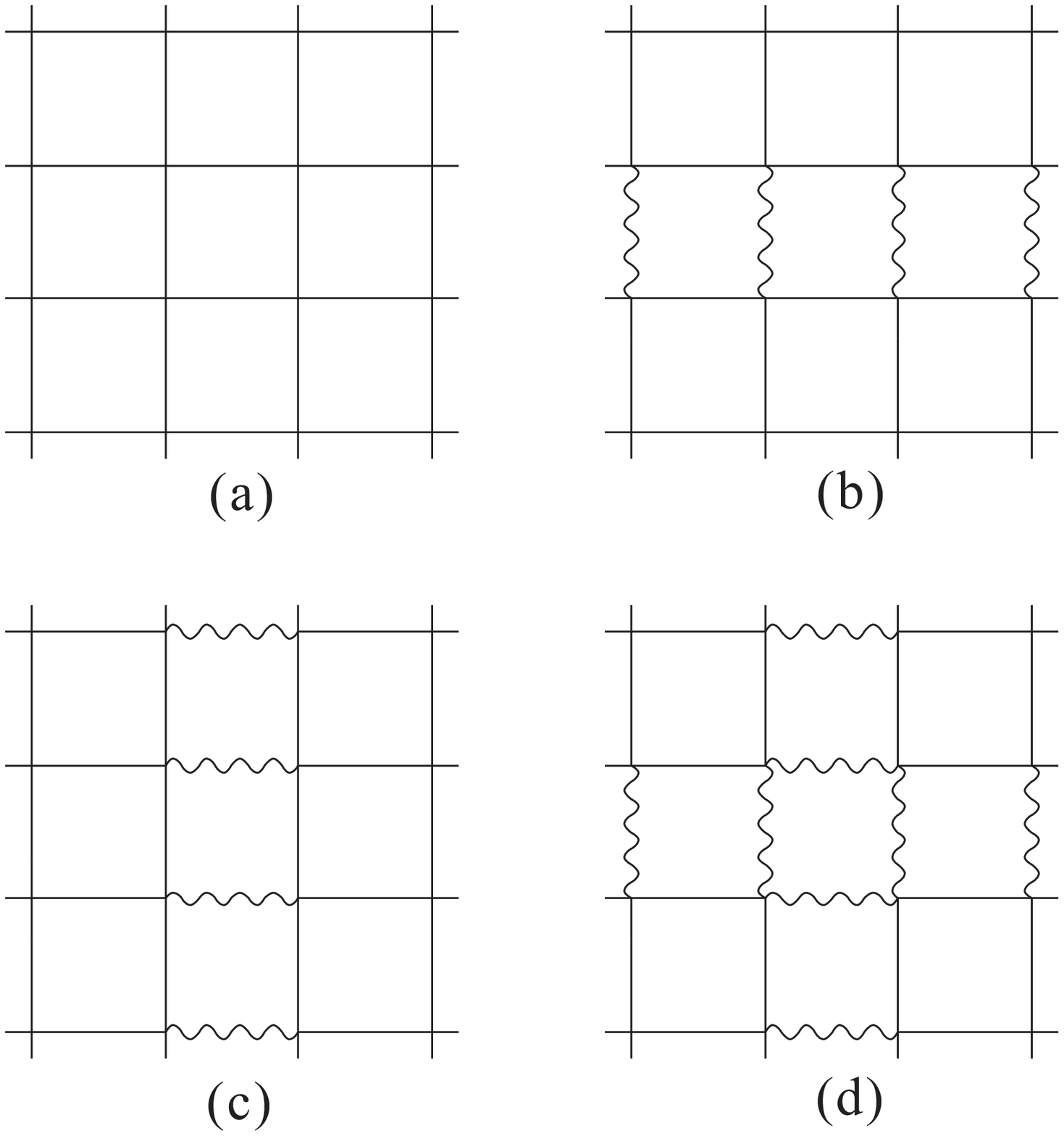}
\end{center}
\caption{Representative cycles of four different classes, $\C_0, \C_1,
\C_2$ and $C_3$ in (a), (b), (c) and (d), respectively.
\label{fig:cycles3}
}
\end{figure}

Cycles are called equivalent if they can be locally deformed into each
other, by repeatedly flipping all spins around some plaquette of the
dual lattice. As illustrated in Fig.~\ref{fig:cycles2}, this
corresponds on the original lattice to applying a product of operators
of the form $X_j = \prod_{\l(j)} X_\l$, where the $\l(j)$ are the four
bonds meeting at site $j$. 
It is then easy to see that there are
exactly four equivalence classes of cycles, denoted $\C_0, \ldots,
\C_3$: $\C_0$ contains all the trivial cycles, which are equivalent to
the empty cycle state 
$|\{\z_\l=1\}\rangle$. $\C_1$ and
$\C_2$ collect cycles equivalent to a single loop winding across the
lattice boundary in the horizontal and vertical directions
respectively, and $\C_3$ those with both a horizontal and a vertical
loop (Fig.~\ref{fig:cycles3}).
In (\ref{codespace}), what is meant
is that for each $|\psi_i\rangle$ the sum runs over all cycle states
$|\z\rangle$ in the equivalence class $\C_i$. This implies in
particular that the $|\psi_i\rangle$ are invariant under the action of
any
of the operators $Z_\pl$ and $X_j$: the $Z_\pl$ leave each cycle state
$|\z\rangle$ invariant, and the $X_j$ only permute cycle states within
an equivalence class,
\begin{equation}
  Z_P|\psi_i\rangle =X_j|\psi_i\rangle =|\psi_i\rangle.
  \label{ZpXj}
\end{equation}

In the large-$N$ limit, the toric code has zero code rate $R$: it
encodes $k=2$ qubits using $n=2N$ qubits, so that $R=k/n=1/N\to
0$. The point of this highly redundant encoding is to allow for the
correction of errors arising from decoherence caused by the
interaction of
the quantum state with its environment. An error introduced in this
way corresponds to a product of Pauli operators $X$, $Y$, $Z$ acting
on some of the quantum spins. Because of $Y=iXZ$, $Y$-errors can be
treated as combinations of $X$- and $Z$-errors.
We write a state with $X$-errors on a set of bonds $\Ss_x$
and $Z$-errors on $\Ss_z$ as
\begin{equation}
 |\tilde{\psi_i} \rangle=
\left(\prod_{b\in \Ss_x}X_b \prod_{b\in\Ss_z}Z_b\,
\right)|\psi_i\rangle.
\label{tilde_psi_i}
\end{equation}
%
If the product of $X_b$ and $Z_b$ on the right-hand side can be
represented
by a product of $X_j$ and $Z_P$ only, then
$|\tilde{\psi_i} \rangle =|\psi_i\rangle$.  
In general, however, this is not the case,
$|\tilde{\psi_i} \rangle \ne |\psi_i\rangle$.

Furthermore, the toric
code is in the general class of Calderbank-Shor-Steane (CSS)
codes~\cite{CSS,CSS2}, for which the $X$- and $Z$-errors can be
treated
separately, without interference between the corresponding error
correction procedures because $[Z_P,X_j]=0$ for any $P$ and $j$
\cite{DKLP}. We can therefore focus in the following
exclusively on $X$-errors, i.e.\ spin flips.  $Z$-errors can be
discussed
separately in the same manner but on the dual lattice \cite{DKLP}.

\subsection{Syndrome and error correction}

We define an {\em error pattern} $\{\tau_\l\}$ such that $\tau_\l=-1$
if an $X$-error has occurred on bond $\l$ (i.e.\ if $\l\in\Ss_x$) and
$\tau_\l=1$ otherwise. To diagnose where errors have occurred, one
measures all the $Z_\pl$. Without corruption all measured values would
be 1 according to (\ref{ZpXj}). The errors give a nontrivial set of
measurement values $\prod_{\l(\pl)} \tau_\l=- 1$ at those plaquettes
around which an odd number of errors have occurred, giving the {\em
syndrome}, the set of plaquettes with measurement value $-1$. It
should be stressed that the syndrome measurement does not cause any
quantum decoherence: From (\ref{tilde_psi_i}), one easily sees -- by
commuting $Z_P$ to the right through the various $X_\l$ and using
(\ref{ZpXj}) -- that even a corrupted state $|\tilde{\psi_i}\rangle$
is an eigenstate of each $Z_P$. Explicitly, $Z_P
|\tilde{\psi_i}\rangle =\pm |\tilde{\psi_i}\rangle$, with the plus
sign when there are an even number of errors around $P$ and the minus
sign otherwise.

\begin{figure}
\begin{center}
\includegraphics[width=0.6\linewidth]{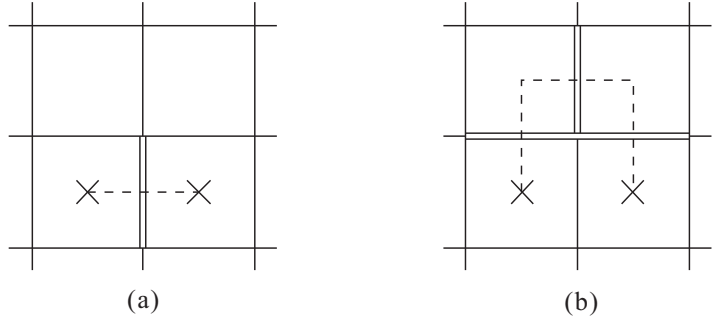}
\end{center}
\caption{Error patterns with the same syndrome (crosses).
Double lines are bonds where $\tau_b=-1$.
Error chains are written in dashed lines.
\label{fig:syndrome}
}
\end{figure}

One can visualize
the error pattern $\tau$ by drawing the duals to the bonds with
$\tau_\l=-1$; these form ``error chains'' ending in ``defects'', i.e.\
plaquettes where the syndrome has detected an error
(Fig.~\ref{fig:syndrome}).
The syndrome measurement is therefore highly ambiguous: any error
chain
$\tau'$ with the same set of defects as $\tau$ gives the same syndrome
as exemplified in Fig.~\ref{fig:syndrome}. The
condition for this is $\prod_{\l(\pl)} \tau_\l =
\prod_{\l(\pl)}\tau'_\l$ for all plaquettes $\pl$ and hence
$\prod_{\l(\pl)} \tau_\l \tau'_\l=1$: the bonds with
$\tau_l\tau'_l=-1$ form a cycle $\tau\tau'$ in the sense defined
above. 

\begin{figure}
\begin{center}
\includegraphics[width=0.6\linewidth]{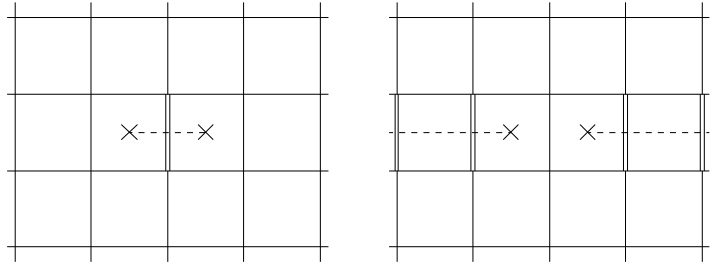}
\end{center}
\caption{Non-equivalent error patterns with the same syndrome.
\label{fig:non_equivalent_patterns}
}
\end{figure}

Now assume we have inferred some error pattern $\tau'$ consistent with
the syndrome (e.g. Fig.\ref{fig:syndrome} (a)),
so that $\tau\tau'$ is a cycle, where $\tau$ is the actual
error pattern (e.g. Fig.\ref{fig:syndrome} (b)). If we correct errors
according to $\tau'$, by applying a spin-flip $X_\l$ to all spins with
$\tau'_\l=-1$, this can be viewed as first correcting the errors
$\tau_\l$
that actually occurred, followed by a series of spin-flips where
$\tau_\l\tau'_\l=-1$ (four double lines in
Fig.~\ref{fig:syndrome} (a) and (b)).
The first stage recovers the uncorrupted code state
$|\psi_i\rangle$. {\em If} $\tau\tau'$ is a {\em trivial} cycle
($\tau\tau'\in
\C_0$), then the second stage corresponds to applying a product of
operators of the form $X_j$. But these leave code states invariant,
see
(\ref{ZpXj}), so
that our error correction was successful. This is a key difference
of this {\em degenerate} quantum code to
classical error correction: we do not need to detect all details of
the error pattern $\tau$, but only its equivalence class (defined as
the set of all $\tau'$ such that $\tau\tau'$ is a trivial cycle).
Error correction according to $\tau'$ will be unsuccessful, on the
other hand, if $\tau$ and $\tau'$ are non-equivalent, i.e.\ if
$\tau\tau'$ is a
{\em nontrivial} cycle like Fig.~\ref{fig:non_equivalent_patterns}:
the second stage from above then mixes up the basis vectors of
different code spaces.

\subsection{Spin-glass representation}

We can now relate the decoding task to a spin glass problem.
The condition that $\tau\tau'$ is a trivial cycle can be written as
$\tau'_{ij}=\tau_{ij} \sigma_i \sigma_j$ if we revert to denoting each
bond
$\l$
by its end points $i$ and $j$. Here the $\sigma_i=\pm 1$ are classical
spin
variables.  Then the second (conceptual) stage of error correction
discussed above would consist of the application of an operator $X_j$
for each site where $\sigma_j=-1$ (the central site in the case of
Fig.~\ref{fig:syndrome}). Assuming that errors were generated
independently and with probability $p$ on each bond $\l$, we assign
an overall probability
\begin{equation}
P(\tau') = \prod_{(ij)} (1-p)^{(1+\tau'_{ij})/2} p^{(1-\tau'_{ij})/2}
 = [p(1-p)]^{n/2}
\exp\left(K_p\sum_{(ij)}\tau'_{ij}\right)
\label{tau_weight}
\end{equation}
to any error chain $\tau'$; here we have defined
$K_p=\frac{1}{2}\ln[(1-p)/p]$. The total probability for error chains
$\tau'$ which lead to successful error correction is therefore
proportional to
\begin{equation}
\Z_0 = \sum_{\tau':\tau\tau'\in\C_0}
\exp\left(K_p\sum_{(ij)} \tau'_{ij}\right) = 
\sum_{\sigma}
\exp\left(K_p\sum_{(ij)} \tau_{ij}\sigma_i \sigma_j\right).
\label{Z_0}
\end{equation}
This is the partition function of an Ising $\pm J$ spin glass with
local interaction strengths $J_{ij}=K_p \tau_{ij}$; the relation
between $K_p$ and $p$ implies that the system is on the so-called
Nishimori line (NL)~\cite{NL,NL2}. If the actual error pattern $\tau$
was
indeed generated according to the assumed probability weight
(\ref{tau_weight}), each bond in (\ref{Z_0}) is ferromagnetic with
probability $1-p$ ($\tau_{ij}=1$) and antiferromagnetic otherwise.

The total probability for error chains $\tau'$ which lead to {\em
faulty} error correction has the same form as (\ref{Z_0}), but with
modified boundary conditions. For example, any $\tau'$ such that
$\tau\tau'\in \C_2$ can be written as $\tau'_{ij}=\tau_{ij} \sigma_i
\sigma_j$ with the convention that all bonds along, say, the left
boundary of the lattice are inverted. This corresponds to the use of
antiperiodic boundary conditions in the left-right direction when
evaluating the spin products $\sigma_i \sigma_j$. The total
probability weight $\Z_2$ for error chains $\tau'$ with $\tau\tau'\in
\C_2$ is therefore a partition function of the form (\ref{Z_0}) with
these modified boundary conditions; in $\Z_0$ we had implicitly
assumed periodic boundary conditions. Similarly $\Z_1$ and $\Z_3$
correspond to the partition functions for antiperiodic boundary
conditions in the top-bottom direction, and in both left-right and
top-bottom directions.

Combining the above results, we conclude that
our total probability of inferring from
the syndrome an error chain $\tau'$ which leads to successful error
correction is $\Z_0/(\Z_0+\Z_1+\Z_2+\Z_3)$ and close to unity as long
as $\Z_0\gg \Z_k$ for $k=1,2,3$. This condition is met if $p$ is small
enough so that we are in the ferromagnetic phase of the spin system
defined by (\ref{Z_0}): the existence of domain boundaries then
implies for the free energies $F_i=-T \ln \Z_i$ that $F_k-F_0$
($k=1,2,3$) is positive and of order $L$, thus $\Z_0\gg \Z_1, \Z_2,
\Z_3$.
In the paramagnetic phase,
on the other hand, $F_k-F_0=\mathcal{O}(1)$ and there will be a
nonzero probability for error correction to fail. In summary, the
toric code can correct errors below an error threshold, i.e.,
in the range $0\leq p\leq p_c$ where the
associated $\pm J$ Ising spin glass on the NL is in its
ferromagnetic phase on the square lattice; $p_c$
is then the location of the multicritical point.
We may therefore be able to learn something about
the spin glass problem from knowledge about the toric code and vice
versa. This is our motivation for the present work.

\section{Simple number counting}
\label{overall}

The argument in the previous section suggests that the numbers of
possible syndromes and equivalence classes of error patterns would
give important measures of
performance of error correction in the toric code.
We therefore discuss this problem in the present and next sections.

Let us first count the total numbers of different syndromes and
equivalence classes, without
specifying the error probability $p$. Since at each plaquette
the syndrome measurement of $Z_\pl$ can give either $+1$ or $-1$, the
total number of syndromes is
\begin{equation}
  D=2^{N-1},
\end{equation}
using that for the square lattice the number of plaquettes is equal to
the number $N=L^2$ of lattice sites. The factor $2^{-1}$ arises
because only an even number of sites with nontrivial syndrome
$Z_\pl=-1$ can exist, as can be seen from the fact that the product of
all $Z_\pl$ is the identity operator on the torus (all $Z_b$ appear
twice from neighbouring plaquettes and $Z_b^2=1$).

The total number of error patterns is $2^n$. And the number of error
patterns in an equivalence class is $2^{N-1}$ because, as explained
above, equivalent error patterns are related by $\tau'_{ij}=\tau_{ij}
\sigma_i \sigma_j$. Each of the $\sigma_i=\pm 1$ can be chosen
independently and gives a different $\tau'$, except for an overall
reversal of the spin configuration which leaves $\tau'$ unchanged and
gives the factor $2^{-1}$. Thus the number of equivalence classes is
\begin{equation}
  C=\frac{2^n}{2^{N-1}}=2^{N+1}
\label{C}
\end{equation}
which is equal to $4D$. This correctly indicates that each syndrome
corresponds to four different equivalence classes of error patterns.
The above argument easily generalizes to other lattices: in general,
$D=2^{\Pl-1}$ where $\Pl$ is the number of plaquettes, while
$C=2^n/2^{N-1}=2^{n-N+1}$. The equality $C=4D$ then follows from
Euler's theorem $n=\Pl+N$.

If we now specify the fraction of errors or error probability $p$,
most of the $2^n$ error patterns are excluded because the number of
errors
these patterns have is different from $np$. Thus the number of
equivalence
classes containing patterns with $np$ errors, $C(p)$, is significantly
smaller than $C$.  Then the number of syndromes $D(=2^{N-1})$ is much
larger than $C(p)$, and we have a sufficient number of syndromes to
specify the equivalence class corresponding to a given syndrome,
$D=C/4 \gg C(p)$. This implies that a simple number counting does not
lead to the critical value $p_c$ by the classical argument that,
beyond $p_c$, the number of errors exceeds that of syndromes
($C(p)>D$) and one cannot identify the error from the syndrome,
leading to unsuccessful error correction.

\section{Bounds on equivalence classes and ground-state energy}
We next proceed to evaluate $C(p)$. Upper and lower bounds are derived
for this quantity, which will further be shown to give a lower
bound on the ground-state energy of the $\pm J$ Ising model.

\subsection{Upper bounds}
\label{upper}

Let us begin by constructing upper bounds.
A trivial upper bound for $C(p)$ is $C$,
\begin{equation}
 C(p)\le C=2^{N+1}\approx 2^N,
\label{ub1}
\end{equation}
where the last expression gives the result to exponential accuracy,
which is all we are normally interested in. To derive another upper
bound, let us temporarily ignore the equivalence of error patterns,
which leads to overcounting. The number of classically distinct error
patterns with a fraction $p$ of errors is
\begin{equation}
  \left(
   \begin{array}{c}
    n\\
    np
   \end{array}
  \right) \approx 2^{nH(p)},
\end{equation}
where $H(p)$ is the binary entropy,
\begin{equation}
 H(p)=-p\log_2 p-(1-p)\log_2 (1-p).
\end{equation}
We therefore have
\begin{equation}
 C(p)\le 2^{nH(p)} \label{ub2}
\end{equation}
because equivalence will reduce the number by grouping errors into
classes.  The two upper bounds (\ref{ub1}) and (\ref{ub2}) cross
each other at the point where $H(p)=\frac{1}{2}$.

An interesting observation is obtained if we continue to ignore the
issue of equivalence and apply the argument for classical codes to the
present problem.  The number of syndromes is $D=2^{N-1}\approx 2^N$.
Thus, if we demand that there exist a sufficient number of syndromes
to distinguish all the errors, we have
\begin{equation}
 2^{nH(p)}<2^N.
   \label{C-bound}
\end{equation}
On the square lattice, $n=2N$, and (\ref{C-bound}) leads to
$H(p)<\frac{1}{2}$ or $p<0.1100$ as a necessary condition for
classical non-degenerate error correction to be successful. This
boundary value $H(p_c)=\frac{1}{2}$ happens to coincide with the
conjecture on the exact location of the multicritical point
(separating the ferromagnetic and paramagnetic phases on the NL) of
the $\pm J$ Ising model on the square lattice, which agrees well with
numerical estimates (see \cite{MKN,NK} and references therein).
Somewhat surprisingly, therefore, a naive argument which {\em ignores}
degeneracy nevertheless seems to give the correct value of the error
threshold for the highly degenerate toric code. \cite{DKLP}.

It is also interesting that the lower bound on the existence of
generic
error-correctable CSS codes, if applied to the toric code, coincides
with the above result as pointed out by DKLP \cite{DKLP}: It is known
that
there exist CSS codes with critical error probability $p_c$ and code
rate
$R=1-2H(p_c)$ in the asymptotic limit of large code size. Since the
toric code has $R\to 0$ asymptotically, one finds
$H(p_c)=\frac{1}{2}$.

\subsection{Lower bound (I)}
\label{lower}

Lower bounds are derived by slightly more elaborate arguments.
%
%
A loose lower bound is
\begin{equation}
  C(p)\ge \frac{2^{nH(p)}}{2^N}.
\end{equation}
The denominator on the right-hand side is the maximum number of error
patterns in an equivalence class, thus leading to a smaller value than
$C(p)$.

A stronger lower bound is given as
\begin{equation}
 C(p)\ge \frac{2^{nH(p_0)}}{2^N},
   \label{C-lb}
\end{equation}
where $p_0$ is a function of $p$ defined by
\begin{equation}
 E_g(p_0)=-n(1-2p)
  \label{Ep0}
\end{equation}
for sufficiently large $N$. Here $E_g(p_0)$ is the ground-state energy
of the $\pm J$ Ising model with a fraction $p_0$ of negative bonds in
a
typical configuration, i.e.\ one where the positive and negative bonds
are randomly distributed.  To prove (\ref{C-lb}), we first
recall from above that the number of error patterns for given $p$,
\begin{equation}
  A(p)=\sum_{\tau}\delta \left( \sum_{(ij)}\tau_{ij}-E_p\right)
    =\left(
     \begin{array}{c}
      n\\
      np
     \end{array}
     \right) \approx 2^{nH(p)},
   \label{Ap}
\end{equation}
where $E_p=n(1-2p)$, is an overcount of the number of equivalence
classes because it ignores equivalence. Each term in (\ref{Ap}) should
be divided by the number of error patterns equivalent to
$\tau$ which contain the same number of errors:
\begin{equation}
  B(p,\tau )=\frac{1}{2}\sum_\sigma 
  \delta \left(\sum_{(ij)}\tau_{ij}\sigma_i \sigma_j-E_p\right).
  \label{Bp}
\end{equation}
The factor $\frac{1}{2}$ comes from overall up-down symmetry (see
before (\ref{C})). We therefore have
\begin{equation}
  C(p)=\sum_{\tau}{}' \frac{\delta \left(
\sum_{(ij)}\tau_{ij}-E_p\right)}
  {B(p,\tau)}.
  \label{C-lb1}
\end{equation}
The prime in the sum indicates that terms with $B(p,\tau)=0$ should be
excluded.  Applying a gauge transformation $\tau_{ij} \to
\tau_{ij}\sigma_i \sigma_j$ and averaging over the gauge variables
$\{\sigma_i\}$ gives, up to an unimportant factor of $\frac{1}{2}$,
 \begin{equation}
   C(p)=\frac{1}{2^N} \sum_{\tau}{}' \frac{\sum_\sigma \delta (\sum
\tau_{ij}\sigma_i\sigma_j-E_p)} 
   {B(p,\tau)}=\frac{1}{2^N}\sum_{\tau}{}' \frac{B(p,\tau)}{B(p,\tau)}
   \equiv \frac{1}{2^N}\sum_\tau \Theta (B(p,\tau)),
   \label{C-lb2}
 \end{equation}
where $\Theta (x)$ is 1 for $x>0$ and 0 for $x=0$. This result is
intuitively clear: the final sum in (\ref{C-lb2}) counts all error
patterns $\tau$ for which an equivalent pattern with a fraction $p$ of
errors exists. Every equivalence class contained in $C(p)$ thus
contributes $2^{N-1}\approx 2^N$ times to the sum, and the prefactor
compensates for this.

The lower bound (\ref{C-lb}) is now obtained by restricting the sum in
(\ref{C-lb2}) to typical error patterns $\tau$ with $np_0$ errors
(i.e.\ $np_0$ negative $\tau_{ij}$'s)
\begin{equation}
  C(p)\ge \frac{1}{2^N} \sum_{\tau (p_0)} \Theta (B(p,\tau)).
\end{equation}
For a typical configuration $\tau$ with $np_0$ negative bonds,
$B(p,\tau)$ is almost always positive because the constraint on the
right-hand side of (\ref{Bp})
 \begin{equation}
   \sum_{(ij)}\tau_{ij}\sigma_i \sigma_j -E_p=0
 \end{equation}
is satisfied by a ground-state configuration of the $\sigma_i$, due to
the definition of $p_0$ in (\ref{Ep0}). We therefore find
\begin{equation}
 C(p)\ge \frac{1}{2^N} \left( 
  \begin{array}{c}
  n\\ np_0
  \end{array}
  \right) \approx 2^{nH(p_0)-N}
\end{equation}
which is the lower bound in (\ref{C-lb}).

\subsection{Lower bound (II)}
\label{lower2}

Another lower bound is obtained by restricting the sum in
(\ref{C-lb1}) to typical bond configurations $\tau$ with $np$ negative
bonds (errors). Then $B(p,\tau)$ is -- in spin glass terms -- the
number of spin configurations with energy $-E_p=-n(1-2p)$ for a
typical
configuration $\tau$ with $np$ negative bonds, the logarithm
of which is the thermodynamic entropy on the NL
\cite{NL,NL2}:
\begin{equation}
 B(p,\tau) =e^{S(p,\tau )}.
\end{equation}
Since $\tau$ is a typical configuration, $S(p,\tau)$ does not depend
on the details of $\tau$ for sufficiently large system size, so we
denote
it as $S(p)$. Thus the
denominator on the right-hand side of (\ref{C-lb1}) can be brought in
front of the sum and we find
\begin{equation}
 C(p)\ge e^{-S(p)} \sum_{\tau(p)} \delta \left(
\sum_{(ij)} \tau_{ij}-E_p\right) \approx 2^{nH(p)-S(p)/\ln 2}.
\end{equation}

\subsection{Behaviour of bounds}
\label{bounds_s}

\begin{figure}[ht]
\begin{center}
\includegraphics[width=0.6\linewidth]{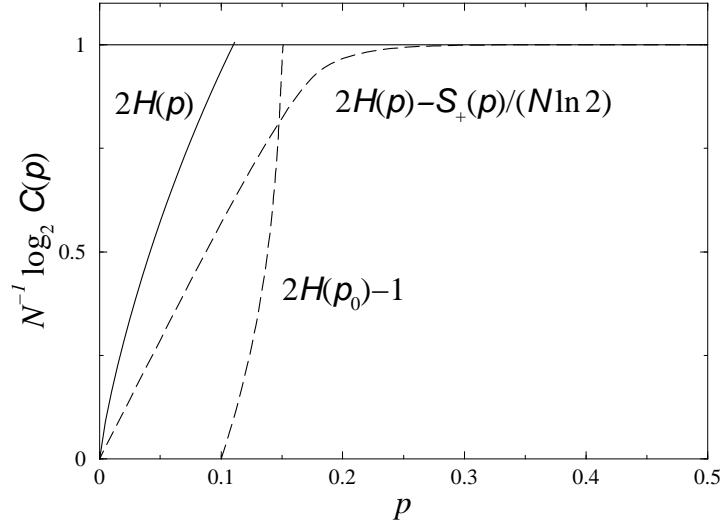}
\end{center}
\caption{Upper and lower bounds of $C(p)$ from (\protect\ref{ineq1}),
plotted as bounds on $N^{-1}\log_2 C(p)$. Solid lines give the upper
bounds, 1 and $2H(p)$, and dashed lines the lower ones,
$2H(p)-S(p)/(N\ln2)$ and $2H(p_0)-1$. Because $S(p)$ is not known
exactly, we replace it by an upper bound $S_+(p)$
from our previous study.~\protect\cite{NS} To sketch $2H(p_0)-1$, an
rough fit was made
to the numerical data in Fig.~\ref{fig:gs} to obtain $p(p_0)$ and from
there the inverse relation $p_0(p)$.
\label{fig:bounds}
}
\end{figure}

Our upper and lower bounds can be summarized as
\begin{equation}
 \max \left( 2^{nH(p)-S(p)/\ln 2},2^{nH(p_0)-N}\right) \le C(p) \le
 \min \left( 2^N, 2^{nH(p)}\right).
  \label{ineq1}
\end{equation}
This result is depicted in Fig.~\ref{fig:bounds}, where we are using
for $S(p)$ an upper bound from our previous work\cite{NS}.
It is seen that $C(p)$
reaches its maximum value $2^N$ at some $p$ below $0.15$ and above
$0.1100$. This is because one of the lower bounds $2^{nH(p_0)-N}$
reaches the upper bound $2^N$ at $p$ close to $0.15$.  The latter
value was obtained by the relation $nH(p_0)-N=N$ (or
$p_0=\frac{1}{2}$) and using the numerical value of the ground-state
energy $E_g(\frac{1}{2})=-1.40N$ (see Fig.~\ref{fig:gs} for this
value).

This saturation of the upper bound at an intermediate value of $p$ is
not
unnatural because $C(p)$ is the number of classes of equivalent error
patterns, with all classes counted with uniform weight. If we instead
give appropriate probability weights to various error patterns, and
thence to equivalence classes, we would reach a smaller value, the
logarithm of which we denote by $S_\pi$. Apart from a trivial factor
of $\ln 2$, this quantity is nothing but the lower bound
$2^{nH(p)-S(p)/\ln 2}$ because the latter was derived by using typical
error configurations,
\begin{equation}
  S_\pi= nH(p)\ln 2-S(p).
  \label{Spi}
\end{equation}
This $S_{\pi}$ is of course smaller than the maximum value of $\ln
C(p)$
and reaches its maximum only at $p=\frac{1}{2}$.

It is instructive to derive (\ref{Spi}) from a different argument.
Consider the probability weight of an error pattern $\tau$ as given in
(\ref{tau_weight}).
%
%
The probability weight of the equivalence class $\pi$ containing this
error
pattern is then 
\begin{equation}
P_\pi = [p(1-p)]^{n/2}\Z_0(p,\tau),
\label{Ppi}
\end{equation}
where $\Z_0(p,\tau)$ is the partition function (\ref{Z_0}) for the
given $p$ (on the NL) and $\tau$. Then the information-theoretical
entropy of the probability distribution $P_\pi$ of equivalence classes
is
\begin{equation}
 S_\pi=-\sum_\pi P_\pi \ln P_\pi,
 \label{Spi2}
\end{equation}
which is equal to the information-theoretical entropy of distribution
of
frustrated plaquettes in the $\pm J$ Ising model \cite{NS, N86}.
Using (\ref{Ppi}) and the relation $F=-T\ln \Z = E-TS$ with
$E=-n(1-2p)$ on the NL, it is possible to reduce this expression to
(\ref{Spi}). To see this, (\ref{Spi2}) is first rewritten using
(\ref{Ppi}) as
\begin{equation}
 S_\pi=-\frac{n}{2} \ln [p(1-p)] +\frac{E}{T}-S(p).
\end{equation}
With $E/T=-n(1-2p)\cdot \frac{1}{2}\ln [(1-p)/p]$ on the NL, we
easily recover (\ref{Spi}).

We can summarize the difference between $\ln C(p)$ and $S_\pi$ as
follows. In $\ln C(p)$ we count (the logarithm of) the total number of
different equivalence
classes $\pi$ that can be obtained for given $p$, i.e.\ that have
$P_\pi>0$: $\ln C(p)$ thus measures the size of the {\em support} of
the
distribution $P_\pi$. On the other hand, $S_\pi$ is the {\em entropy}
of the distribution, and the fact that it is smaller than $\ln C(p)$
indicates that the distribution is strongly (i.e. exponentially
narrowly) peaked rather than uniformly spread over its support.

\subsection{Lower bound on the ground-state energy}
\label{ground_state}

\begin{figure}
\begin{center}
\includegraphics[width=0.6\linewidth]{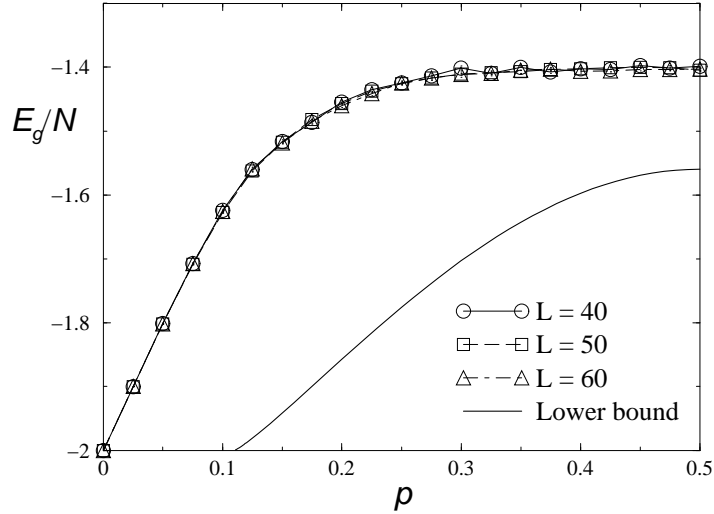}
\end{center}
\caption{Ground-state energy of the two-dimensional $\pm J$ spin
glass. Shown is the lower bound resulting
from~(\protect\ref{Eg_bound}), compared with numerical results for
linear system sizes $L=40$, 50 and 60. The latter were obtained by
averaging over 10, 15 and 5 numerical realizations of the disorder,
respectively; error bars are smaller than the symbol sizes. All
numerical calculations were run on the spin glass ground state
server~\protect\cite{sgs}.
\label{fig:gs}
}
\end{figure}

The inequality (\ref{ineq1}) implies the relation
\begin{equation}
  nH(p)\ge nH(p_0)-N.
\label{Eg_bound}
 \end{equation}
If we regard $p$ as a function of $p_0$ through the definition
$E_g(p_0)=-n(1-2p)$, then the above inequality gives a lower bound on
$p(p_0)$, or equivalently a lower bound on $E_g(p_0)$.  The result is
shown in Fig.~\ref{fig:gs} together with numerical estimates of $E_g$.
Our lower bound is not particularly tight numerically, e.g.\
$E_g(p)/N\ge -1.56$ at $p=\frac{1}{2}$ whereas numerically $E_g(p)/N$
is around $-1.40$.  However, this result has
non-trivial significance because it is, as far as we know, the first
analytical lower bound on the ground-state energy of the
two-dimensional $\pm J$ Ising model with general $p$.

\section{Summary and discussion}
\label{conclusion}

We have derived upper and lower bounds on the number of equivalence
classes of error patterns in the toric code. It has been shown that
this number saturates its upper bound at an intermediate value of
the error probability where we expect no singularities in physical
quantities.  This apparently non-conventional behaviour has been
explained by noting that the number of {\em typically realized}
equivalence
classes, which is relevant for physical quantities, is significantly
smaller than the
number of equivalence classes with uniform weights given to all the
cases. The logarithm of the former number has been shown to be equal
to the information-theoretical entropy of the probability of error
classes, which is further related to the thermodynamic entropy on the
NL and
therefore has a singularity (albeit a weak one) at the multicritical
point.

One of the upper bounds was compared with a lower bound, the latter
involving the ground-state energy of the $\pm J$ Ising model, leading
to a lower bound on the ground-state energy of this spin glass model.
Although the resulting value of the lower bound is not necessarily
impressive numerically, it is interesting that bounds on the number
of equivalence classes of the toric code lead in a simple manner to a
bound on the ground-state energy of a spin glass. The correspondence
between the two problems was proposed by DKLP~\cite{DKLP}, and we have
exploited it here to derive an explicit result on a physical quantity.

The present work would serve as a starting point for further
developments based on the correspondence of two completely different
problems.  For example, an improved upper bound for $C(p)$ will lead
to a better lower bound on the ground-state energy.  Improvements of
the lower bound of $C(p)$ may also be possible, but one should
remember that such a result may not lead to an improved bound on the
ground-state energy unless the obtained lower bound of $C(p)$ is
related to the latter quantity.

\section*{Acknowledgement}
We thank John Preskill for very useful discussions and comments on
equivalence class in the toric code. We
are also grateful to Thomas Lange for technical support with the
simulations on the spin glass ground state server.
The present work has been supported by
the Anglo-Japanese Collaboration Programme by the Japan Society for
the Promotion of Science and The Royal Society. One of the authors
(HN) was also supported by the MEXT Grant-in-Aid for Scientific
Research on Priority Area `Statistical-Mechanical Approach to
Information Processing' and by the MEXT 21st Century COE Programme
`Nanometer-Scale Quantum Physics' at Department of Physics, Tokyo
Institute of Technology. PS gratefully acknowledges the hospitality of
the Tokyo Institute of Technology, where this work was initiated.

\end{document}